\documentclass[aps,pre,twocolumn,showpacs]{revtex4}
\usepackage{graphicx}
\usepackage{amsmath}
\usepackage{amsfonts}
\usepackage{amssymb}

\DeclareMathOperator{\erf}{erf}

\begin{document}

\title{Theoretical description of spherically confined strongly correlated
Yukawa plasmas}
\author{H. Bruhn, H. K\"ahlert, T. Ott}  \author{M. Bonitz} \email{bonitz@physik.uni-kiel.de}
\affiliation{Institut f\"ur Theoretische Physik und Astrophysik,
Christian-Albrechts-Universit\"at zu Kiel, D-24098, Germany}
\author{J. Wrighton and J. W. Dufty}
\affiliation{Department of Physics, University of Florida, Gainesville, Florida 32611, USA}
\date{\today}

\begin{abstract}
A theoretical description of the radial density
profile for charged particles with Yukawa interaction in a harmonic trap is
described. At strong Coulomb coupling shell structure is observed in both
computer simulations and experiments. Correlations responsible for such
shell structure are described here using a recently developed model based in
density functional theory. A wide range of particle number, Coulomb
coupling, and screening lengths is considered within the fluid phase. A
hypernetted chain approximation shows the formation of shell structure, but
fails to give quantitative agreement with Monte Carlo simulation results at
strong coupling. Significantly better agreement is obtained within the
hypernetted chain structure using a renormalized coupling constant,
representing bridge function corrections.
\end{abstract}

\maketitle










\section{Introduction}

\label{intro} Spatially confined charged particles have attracted
growing interest. Examples include electrons in quantum dots
\cite{7}, ions in Penning and Paul traps \cite{3,4} and the
mesoscopic charges of dusty plasmas \cite{c,d}. In particular,
three-dimensional classical spherical plasmas have been produced in
ion systems \cite{Gilbert} and more recently in dusty plasmas
\cite{5}. The structural and dynamic properties of these systems
continue to attract the interest of many groups in various fields; e.g., 
\cite{apolinario11, cioslowski11, kaehlert11}.

At sufficiently strong coupling these systems
form concentric shells which are well reproduced by simulations, c.f.  \cite{rafac_pnas91,ludwig05,e} and references therein.
The objective here is to provide a theoretical analysis to complement these results from
simulation and experiments, for a better physical understanding of the
underlying mechanisms.  For
harmonically confined particles with Coulomb interaction such a theory of
shell formation as a function of temperature (inverse coupling) was derived
recently using classical density functional theory (DFT) \cite{wrighton2009,a8,wrighton2010}.
However, a special property of dusty
plasmas is the screening of the pair interaction. The theoretical
description is extended here to describe spherically trapped strongly correlated particles with
Yukawa interaction for such dusty plasmas.

The state conditions are specified by three dimensionless parameters:
particle number $N$, coupling constant $\Gamma $ (defined below), and
dimensionless screening parameter $\kappa ^{\ast }$. The ranges of values
considered are $15<N<500$, $0<\Gamma <100$, and $0<\kappa ^{\ast }\le 1$. The
primary focus here is on shell formation as a
function of these parameters. Only the equilibrium fluid phase is considered
(rotational invariance) so that shell structure is reflected in the radial
dependence of the density of confined charges. The average density is
defined as a multi-dimensional configuration integral in the canonical
ensemble, which can be evaluated by Monte Carlo simulation. New simulations
are provided here as a means to determine the accuracy of theoretical
approximations. The system, dimensionless units, and adaptation of the
hypernetted chain (HNC) theory introduced in  \cite{a8} for
confined Coulomb charges are described in Section \ref{MCS}.  The density
profiles from the HNC approximation are compared to simulations in Section %
\ref{HNCA}. It is found that the formation of shells, as well as their location and
populations, is well described by the HNC approximation, but the shell
maxima and widths show large discrepancies for $\Gamma >10$ and the errors
increase with increasing $\kappa ^{\ast }$. The primary qualitative effect
of screening is to shift the shells toward the center and decrease the
overall volume. An ``adjusted'' hypernetted chain approximation (AHNC) is
considered in Section \ref{AdHNC}. This is based on a model for the bridge
function corrections to HNC first introduced by Ng \cite{NG} for the pair
distribution function of a Coulomb one component plasma (OCP). It has the
property of preserving the form of the HNC equations with only a
renormalization of $\Gamma $ to some larger effective value. It is
shown that the same method applies to the Yukawa OCP for accurate pair
correlations even at very strong coupling, and the approach is then applied to
the bridge corrections to the equation for the density profile. An optimized
renormalization for the density leads to excellent results for the Coulomb
case (e.g., $\Gamma \leq 100,N\leq 500$).
For Yukawa systems this approach is very useful as well for the radial 
distribution function, but it is
somewhat more limited for the density profile at larger values
 of  $\kappa ^{\ast }$ and $N$.

\section{Theory and Simulation}
\subsection{Model and units}

\label{MCS}The system is comprised of $N$ identical charges interacting
pairwise via a Yukawa potential, confined by a harmonic potential centered
at the origin. The Hamiltonian is
\begin{equation}
H=\sum_{i=1}^{N}\left( \frac{1}{2}mv_{i}^{2}+\frac{1}{2}m\omega
^{2}r_{i}^{2}\right) +\frac{1}{2}\sum_{i\neq j=1}^{N}V(r_{ij}).  \label{2.1}
\end{equation}%
Here $m$ is the mass, $\omega $ is the angular frequency measuring the
strength of the confinement, and $\mathbf{r}_{i}$,$\mathbf{v}_{i}$ are the
position and velocity of charge $i$. The Yukawa interaction is
\begin{equation}
V(r_{ij})=q^{2}\frac{e^{-\kappa r_{ij}}}{r_{ij}},  \label{2.2}
\end{equation}%
where $r_{ij}\equiv \left\vert \mathbf{r}_{i}-\mathbf{r}_{j}\right\vert $, $%
q $ is the particle charge, and $\kappa $ is an inverse screening length.
The physical origin of this screening length is described elsewhere \cite{d} and
will not be discussed here. The primary property of interest here is the
local density of charges in the trap at equilibrium. For the classical
canonical ensemble its dimensionless form is given by%
\begin{equation}
n^{\ast }(r_{1}^{\ast })\equiv n(r_{1})r_{0}^{3}=
N\frac{\int d\mathbf{r}%
_{2}^{\ast }\dots d\mathbf{r}_{N}^{\ast }e^{-V^{\ast }(\mathbf{r}_{1}^{\ast },..,%
\mathbf{r}_{N}^{\ast })}}{\int d\mathbf{r}_{1}^{\ast } \dots d\mathbf{r}%
_{N}^{\ast }e^{-V^{\ast }(\mathbf{r}_{1}^{\ast },..,\mathbf{r}_{N}^{\ast })}}%
,  \label{2.3}
\end{equation}%
with%
\begin{eqnarray}\label{2.4}
V^{\ast }(\mathbf{r}_{1}^{\ast },..,\mathbf{r}_{N}^{\ast }) & \equiv & \beta V(%
\mathbf{r}_{1},..,\mathbf{r}_{N})= \\
&=& \Gamma \left[ \frac{m\omega ^{2}r_{0}^{3}}{%
2q^{2}}\sum_{i=1}^{N}r_{i}^{\ast 2}+\frac{1}{2}\sum_{i\neq j=1}^{N}\frac{%
e^{-\kappa ^{\ast }r_{ij}^{\ast }}}{r_{ij}^{\ast }}\right] .
\nonumber
\end{eqnarray}%
Here, $\mathbf{r}_{i}^{\ast }=\mathbf{r}_{i}/r_{0}$, $\beta =1/k_{B}T$ is
the inverse temperature, $\Gamma =\beta q^{2}/r_{0}$ is the Coulomb coupling
constant, and $\kappa ^{\ast }=\kappa r_{0}$. The usual choice for the
length scale $r_{0}$ is the ion sphere radius, or mean distance between
charges, given by
\begin{equation}
4\pi r_{0}^{3}\overline{n}/3=1,  \label{2.4a}
\end{equation}%
where $\overline{n}$ is a characteristic spatially averaged density to be
chosen for convenience. Here it is chosen to simplify the Hamiltonian by the
condition
\begin{equation}
\frac{m\omega ^{2}r_{0}^{3}}{2q^{2}}=\frac{1}{2},\hspace{0.2in}\text{or \ \ }%
\overline{n}=\frac{3m\omega ^{2}}{4\pi q^{2}}.  \label{2.4b}
\end{equation}%
This is not the average density for the Yukawa particles in the trap $%
\overline{n}_{T}=N/V_{T}$, where the volume $V_{T}=4\pi R_{T}^{3}/3$ is
defined by the maximum radius $R_{T}$ at which the force on a charge due to
the trap is equal to that of all other charges
\begin{equation}
m\omega ^{2}R_{T}=q^{2}\int d\mathbf{r}^{\prime }\frac{e^{-\kappa \left\vert
\mathbf{R}_{T}-\mathbf{r}^{\prime }\right\vert }}{\left\vert \mathbf{R}_{T}-%
\mathbf{r}^{\prime }\right\vert ^{2}} \left(1+ \kappa\left\vert
\mathbf{R}_{T}-\mathbf{r}^{\prime }\right\vert \right)   n_{T}(r^{\prime }).  \label{2.5}
\end{equation}%
It follows that $\overline{n}_{T}=\overline{n}$ for $\kappa =0$ and $%
\overline{n}_{T}>\overline{n}$ \ for $\kappa >0$. The solution to (\ref{2.5}%
) is discussed further below.

The dimensionless trap potential energy is now a function of two
dimensionless parameters, the Coulomb coupling strength $\Gamma $ and the
screening parameter $\kappa ^{\ast }$
\begin{equation}
V^{\ast }(\mathbf{r}_{1}^{\ast },..,\mathbf{r}_{N}^{\ast })=\Gamma \frac{1}{2%
}\left[ \sum_{i=1}^{N}r_{i}^{\ast 2}+\sum_{i\neq j}^{N}\frac{e^{-\kappa
^{\ast }\left\vert \mathbf{r}_{i}^{\ast }-\mathbf{r}_{j}^{\ast }\right\vert }%
}{\left\vert \mathbf{r}_{i}^{\ast }-\mathbf{r}_{j}^{\ast }\right\vert }%
\right] ,  \label{2.6}
\end{equation}
and the dimensionless density profile $n^{\ast }(r^{\ast })$ can be
obtained numerically from a Metropolis Monte Carlo simulation for
given $\Gamma ,\kappa ^{\ast },$ and $N$.

\subsection{Theory and approximations}

A formal representation of the average density profile was developed within
density functional theory in reference \cite{a8}. That analysis
applies here as well, with only the replacement of the Coulomb potential by
the Yukawa potential. First, the density is represented in terms of a
dimensionless effective potential $U^{\ast }\left( r^{\ast }\right) $
\begin{equation}
n^{\ast }(r^{\ast })\equiv \overline{N}\frac{e^{-\Gamma U^{\ast }\left(
r^{\ast }\right) }}{4\pi \int_{0}^{\infty }dr^{\ast }r^{\ast 2}e^{-\Gamma
U^{\ast }\left( r^{\ast }\right) }}.  \label{3.1}
\end{equation}%
Here $\overline{N}$ denotes the average number of particles in the trap,
since the theory is formulated in the grand canonical ensemble. The
effective potential obeys the equation%
\begin{equation}
U^{\ast }\left( r^{\ast }\right) =\frac{1}{2}r^{\ast 2}+\overline{N}\frac{%
\int d\mathbf{r}^{\ast \prime }e^{-\Gamma U^{\ast }\left( r^{\ast \prime
}\right) }\overline{c}(|\mathbf{r}^{\ast }-\mathbf{r}^{\ast \prime }|)}{\int
d\mathbf{r}^{\ast \prime }e^{-\Gamma U^{\ast }\left( r^{\ast \prime }\right)
}}+B\left( r^{\ast }\right) .  \label{3.2}
\end{equation}%
The function $\overline{c}(r^{\ast })$ is proportional to the direct
correlation function for a uniform one component plasma (OCP) of Yukawa
charges%
\begin{equation}
\overline{c}(r^{\ast })=-\frac{1}{\Gamma }c_{\text{OCP}}(r^{\ast }),
\label{3.3}
\end{equation}%
which is related to the OCP radial distribution
function $g_{\text{OCP}}(r^{\ast })$ by the Ornstein-Zernike equation%
\begin{eqnarray}\label{3.4}
&& g_{\text{OCP}}(r^{\ast })-1 = c_{\text{OCP}}(r^{\ast })+\\\nonumber
&+& \overline{n}_{\text{OCP%
}}^{\ast }\int d\mathbf{r}^{\ast \prime }\left[ g_{\text{OCP}}(r^{\ast
\prime })-1\right] c_{\text{OCP}}(\left\vert \mathbf{r}^{\ast }-\mathbf{r}%
^{\ast \prime }\right\vert ).
\end{eqnarray}%
Finally, $g_{\text{OCP}}(r^{\ast })$ is determined from the equation%
\begin{eqnarray} \label{3.5}
&& \ln {g_{\text{OCP}}(r^{\ast })}=-\Gamma \frac{e^{-\kappa ^{\ast }r^{\ast }}}{%
r^{\ast }}+ \\\nonumber
&+& \overline{n}_{\text{OCP}}^{\ast }\int d\mathbf{r}^{\ast \prime
}\left[ g_{\text{OCP}}(r^{\ast \prime })-1\right] c_{\text{OCP}}\left( |%
\mathbf{r}^{\ast }-\mathbf{r}^{\ast \prime }|\right) -\\\nonumber
&-& \Gamma B_{\text{OCP}}\left( \mathbf{r}^{\ast }\right) .
\end{eqnarray}

The second term of (\ref{3.2}) describes the effect of correlations among
particles in the trap in terms of the corresponding \ correlations among
particles in the uniform OCP. The last term $B\left( r^{\ast }\right) $
corrects this approximate treatment of correlations and is known as a bridge
function. Similarly, $B_{\text{OCP}}\left( \mathbf{r}^{\ast }\right) $ is
the bridge function for ${g_{\text{OCP}}(r^{\ast })}$ \cite{ToSL}. To
optimize this contribution of OCP correlations, the density of the trap is
matched to that of the OCP%
\begin{equation}
\overline{n}_{\text{OCP}}^{\ast }=\overline{n}_{\text{T}}^{\ast }.
\label{3.6}
\end{equation}
For given $\overline{N}$ the trap density is fixed by the volume of the
trap, whose radius $R_{T}$ must be calculated from (\ref{2.5}). An
approximate evaluation for the ground state has been discussed elsewhere \cite{14}, with the result
that it is the unique positive, real solution to
\begin{eqnarray}
-(1+\kappa ^{\ast }R^{\ast })(\overline{N}-1)+R^{\ast 3}+\kappa ^{\ast }R^{\ast 4}\nonumber\\
+\frac{6}{15}%
\kappa ^{\ast 2}R^{\ast 5}+\frac{1}{15}\kappa ^{\ast 3}R^{\ast 6}=0.
\label{3.7}
\end{eqnarray}%
In all of the following, $\overline{n}_{\text{T}}^{\ast }$ is determined in
this way for each $\kappa ^{\ast }$.

The above equations (\ref{3.1}) - (\ref{3.5}) are still exact, but require
specification of the bridge functions. The simplest approximation is the
neglect of the bridge functions, leading to the hypernetted chain
approximation (HNC)%
\begin{equation}
U_{\text{HNC}}^{\ast }\left( r^{\ast }\right) =\frac{1}{2}r^{\ast 2}+%
\overline{N}\frac{\int d\mathbf{r}^{\ast \prime }e^{-\Gamma U^{\ast }_{\text{HNC}} \left(
r^{\ast \prime }\right) }\overline{c}_{\text{HNC}}(|\mathbf{r}^{\ast }-%
\mathbf{r}^{\ast \prime }|)}{\int d\mathbf{r}^{\ast \prime }e^{-\Gamma
U^{\ast }_{\text{HNC}}\left( r^{\ast \prime }\right) }},  \label{3.8}
\end{equation}%
\begin{eqnarray}\label{3.9}
&&\ln {g_{\text{HNC}}(r^{\ast })}=-\Gamma \frac{e^{-\kappa ^{\ast }r^{\ast }}}{%
r^{\ast }}+ \\\nonumber
&&\quad + \overline{n}_{\text{T}}^{\ast }\int d\mathbf{r}^{\ast \prime
}\left[ g_{\text{HNC}}(r^{\ast \prime })-1\right] c_{\text{HNC}}\left( |%
\mathbf{r}^{\ast }-\mathbf{r}^{\ast \prime }|\right) ,
\end{eqnarray}%
\begin{eqnarray}\label{3.10}
&&g_{\text{HNC}}(r^{\ast })-1=c_{\text{HNC}}(r^{\ast })+ \\\nonumber
&&\quad + \overline{n}_{\text{T}%
}^{\ast }\int d\mathbf{r}^{\ast \prime }\left( g_{\text{HNC}}(r^{\ast \prime
})-1\right) c_{\text{HNC}}(\left\vert \mathbf{r}^{\ast }-\mathbf{r}^{\ast
\prime }\right\vert ).
\end{eqnarray}%
This is a closed set of equations for $U_{\text{HNC}}^{\ast },{g_{\text{HNC}%
}(r^{\ast })},$ and $\overline{c}_{\text{HNC}}$. Note that the determination
of ${g_{\text{HNC}}(r^{\ast })}$ and $\overline{c}_{\text{HNC}}$ is
independent of the trap density calculation.
\begin{figure}
\includegraphics[scale=0.7]{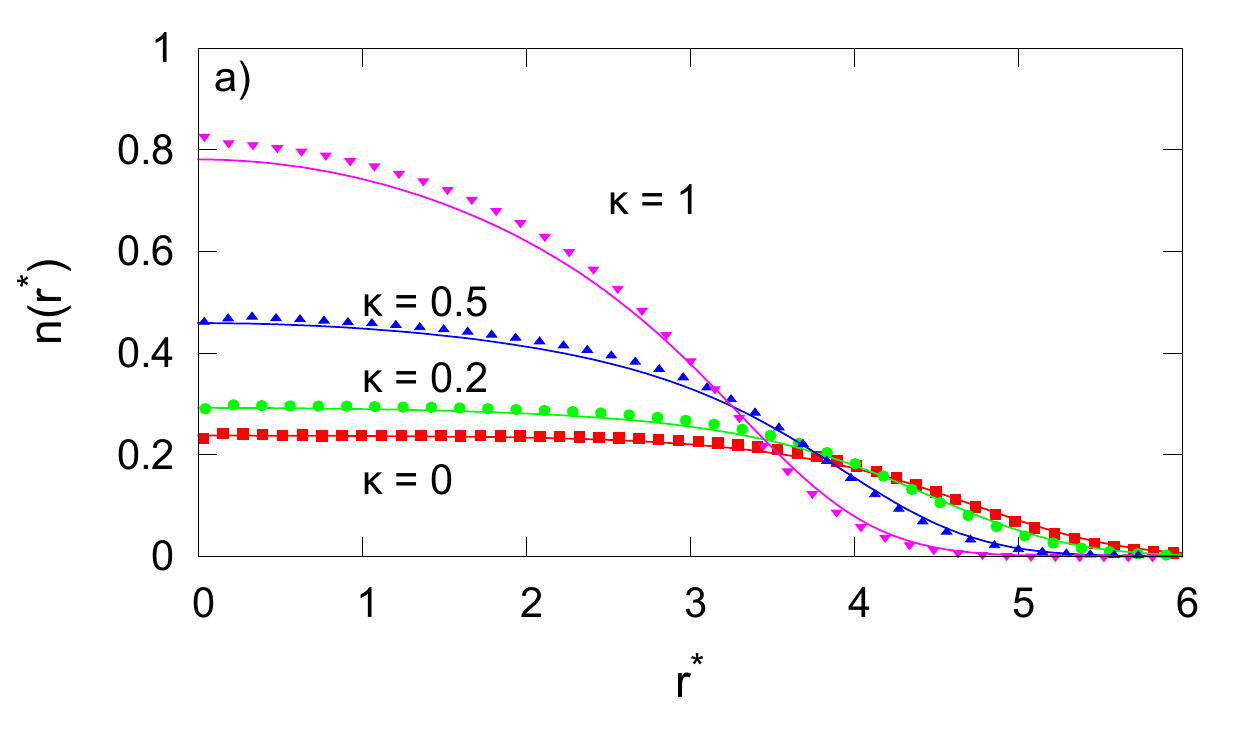}
\hfill
\includegraphics[scale=0.7]{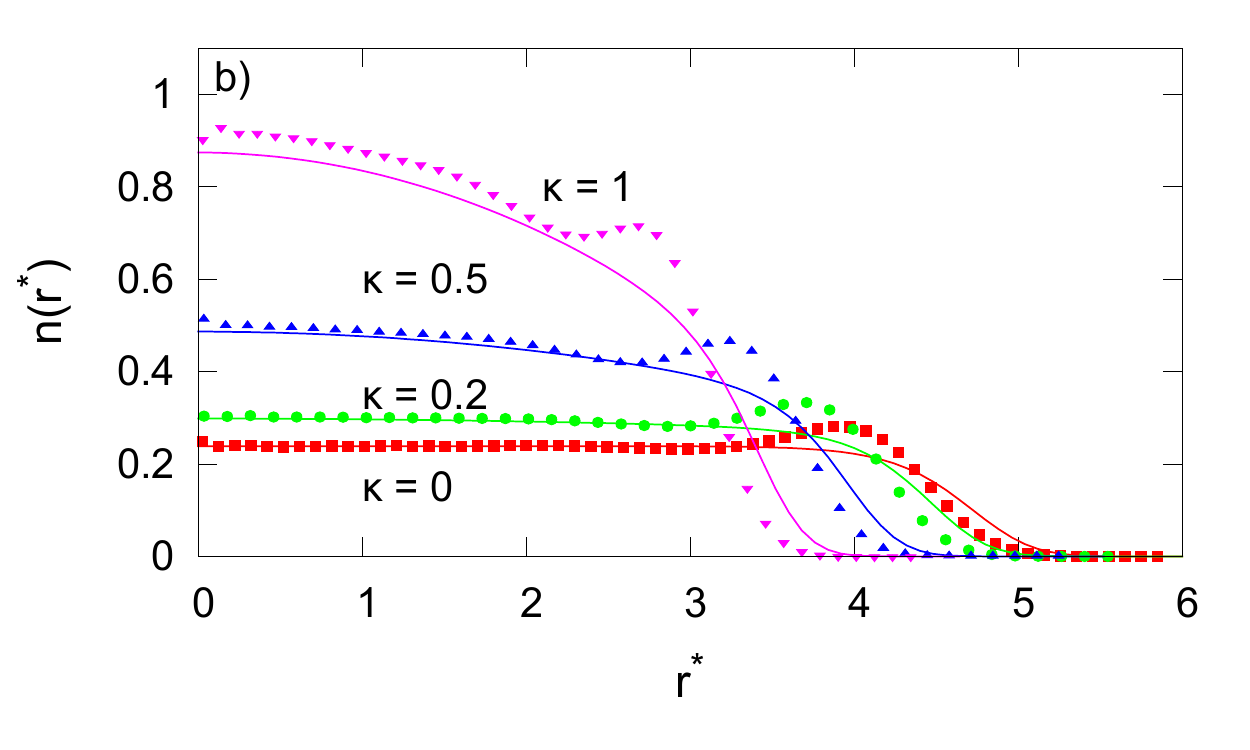}
\caption{Mean field results for the density profile (lines) for a Coulomb and Yukawa OCP in a spherical trap for (a) $\Gamma=1$ and (b) $\Gamma=5$ for $N=100$ compared with Monte Carlo simulations (symbols).}
\label{fig1}
\end{figure}

It is well known that the HNC approximation for the OCP properties is a good
approximation except for strong coupling where the bridge function $B_{\text{%
OCP}}$ becomes important. However, the results below for the trap density
show that the trap bridge function can be important even at moderate
coupling. It is therefore necessary to go beyond HNC and find an
approximation for the bridge functions. This is described below.

\section{Results: HNC Approximation}\label{HNCA}

\begin{figure}[h]
\includegraphics[scale=0.7]{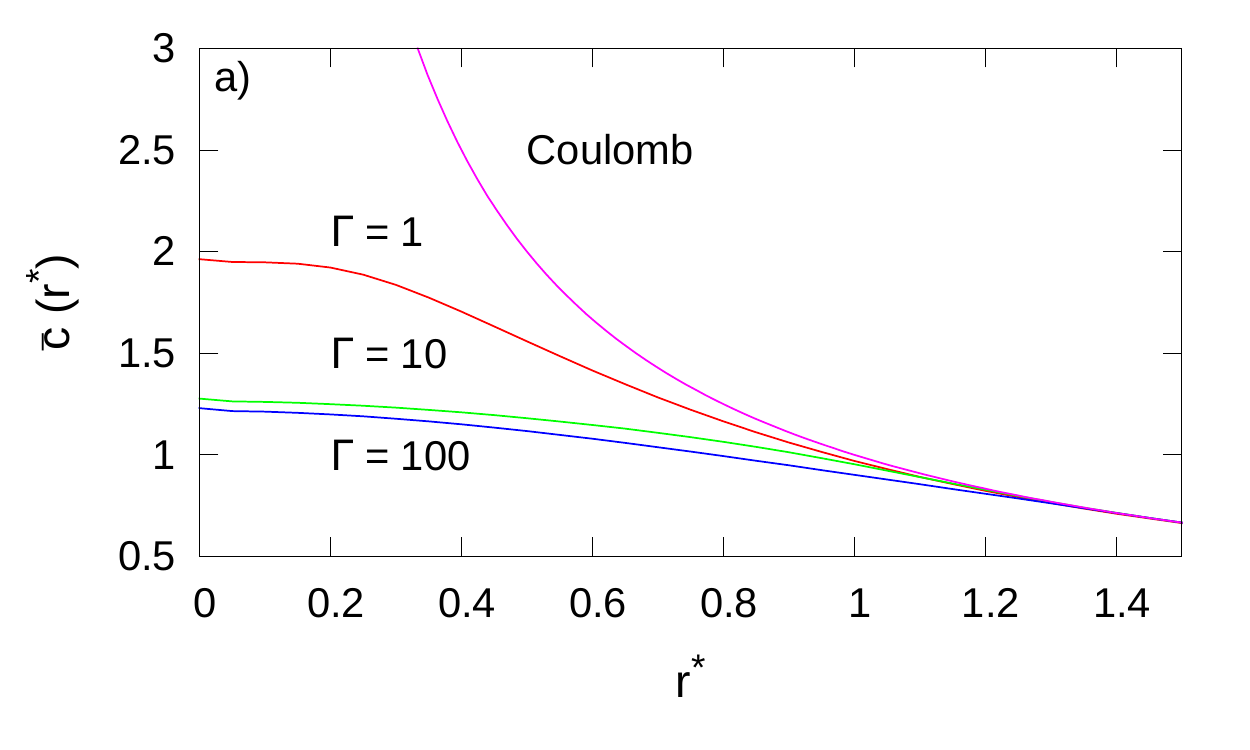} \hfill \includegraphics[scale=0.7]{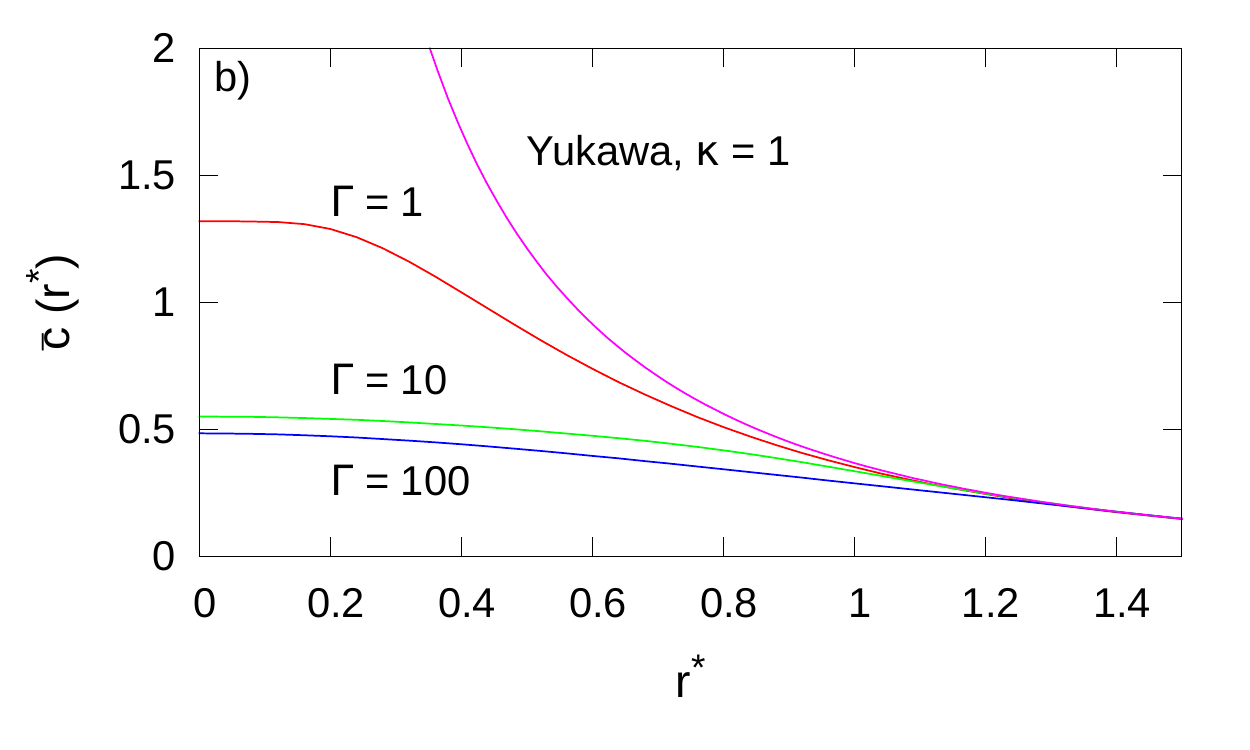}
\caption{Various direct correlation functions for (a) Coulomb interaction and (b) Yukawa interaction with $\protect\kappa^{\ast }=1$.  The top curve is the mean field value.}
\label{fig6}
\end{figure}
Correlations in the HNC approximation are described by $%
\overline{c}_{\text{HNC}}$. For weak coupling, $\Gamma <1$, $\overline{c}_{%
\text{HNC}}\rightarrow e^{-\kappa ^{\ast }r^{\ast }}/r^{\ast }$. This is the
``mean field'' limit. Fig. \ref{fig1} shows a comparison of this mean field
description with Monte Carlo simulation results at moderate coupling, $\Gamma =1$ and $5$ for
several values of $\kappa ^{\ast }$. As might be expected, there is reasonable agreement at $%
\Gamma =1$, but emergence of an outer shell is evident at $\Gamma = 5$, which cannot be
reproduced by the mean field theory. Evidently, here it is necessary to calculate
the correlations of $\overline{c}_{\text{HNC}}$ through the full coupled set
of equations (\ref{3.8}) - (\ref{3.10}).

\begin{figure}[h]
\includegraphics[scale=0.7]{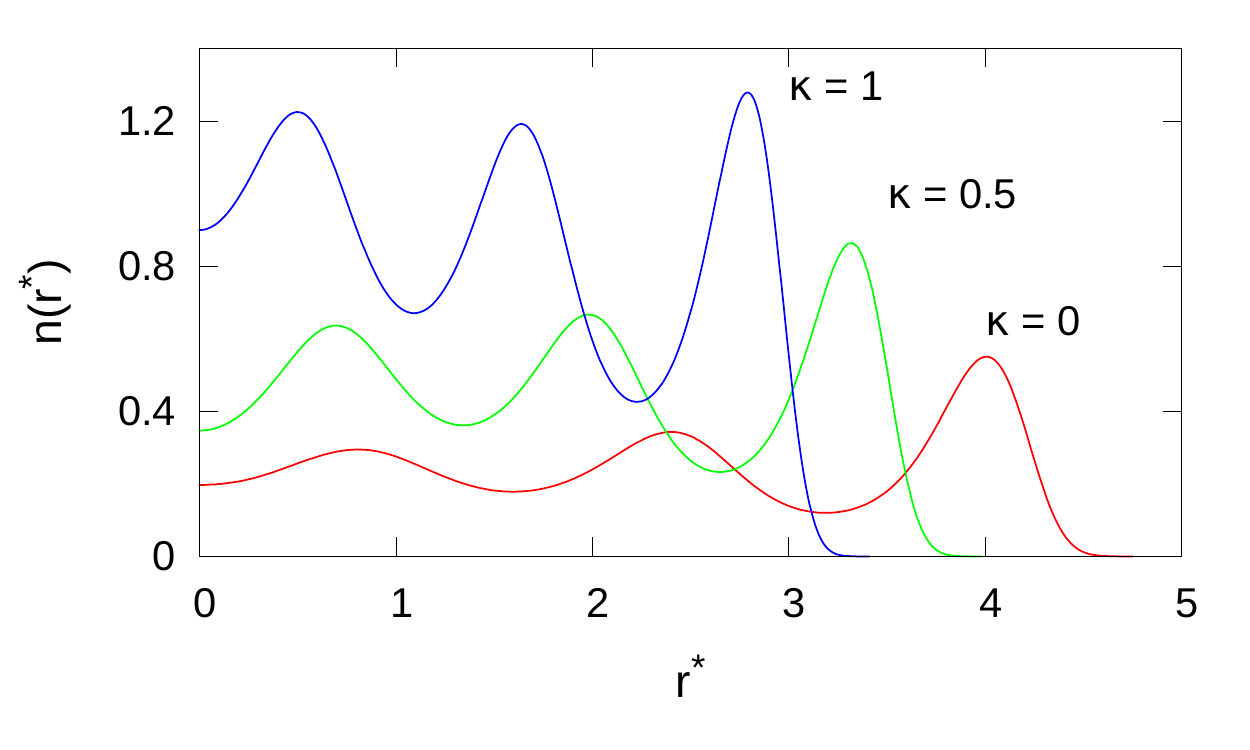}
\caption{HNC density profile for a Yukawa system with various $\protect\kappa^{\ast } $ at $\Gamma =50$ and $N=100$. }
\label{fig9}
\end{figure}

Figures \ref{fig6}a) and \ref{fig6}b) show $\overline{c}_{\text{HNC}}$ as a function of $%
r^{\ast }$ for $\kappa ^{\ast }=0$ and $1$. In both cases the deviation from
the mean field limit increases for stronger coupling, creating a
``correlation hole'' for $r^{\ast }<1$. The effects of these correlations on
the trap density profile are illustrated for several values of the screening
parameter $\kappa ^{\ast }$ in Fig. \ref{fig9} at $\Gamma =50,N=100$. It is
seen that increased screening tends to compress the system \cite{e} and enhance the
shell structure.

%
%

\begin{figure}[tbp]
\includegraphics[scale=0.6]{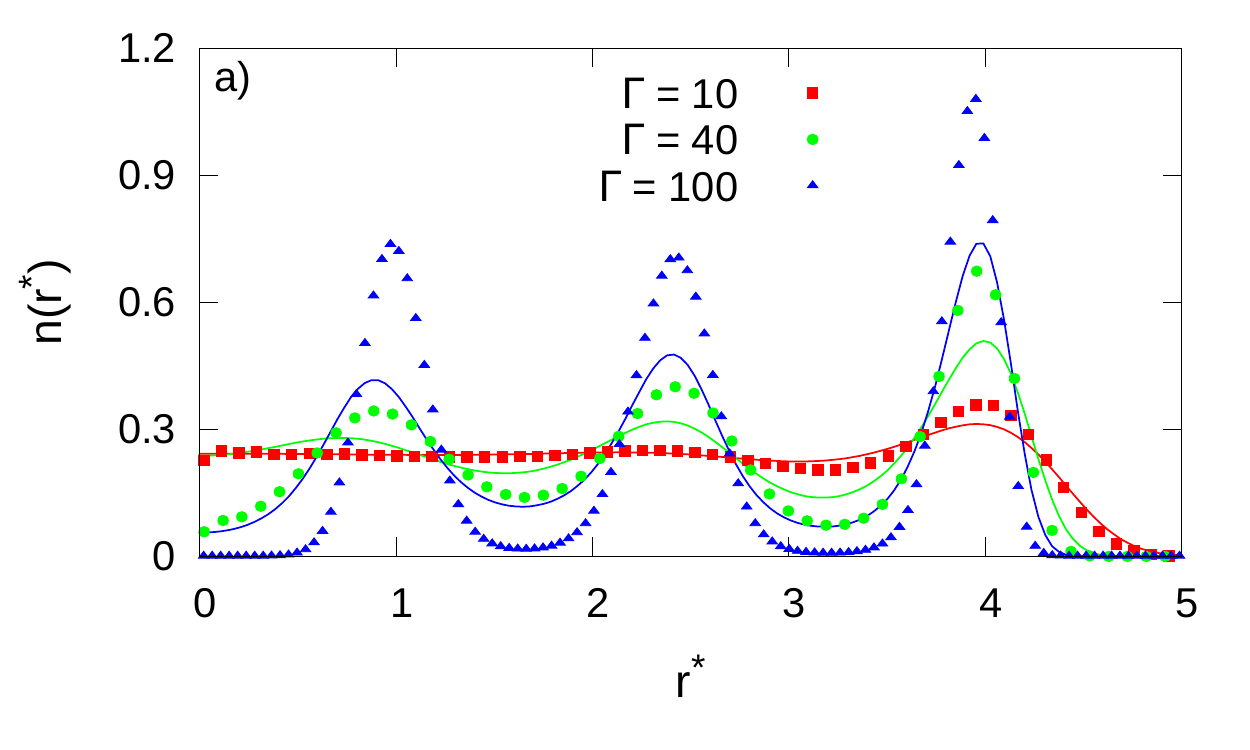}
\includegraphics[scale=0.6]{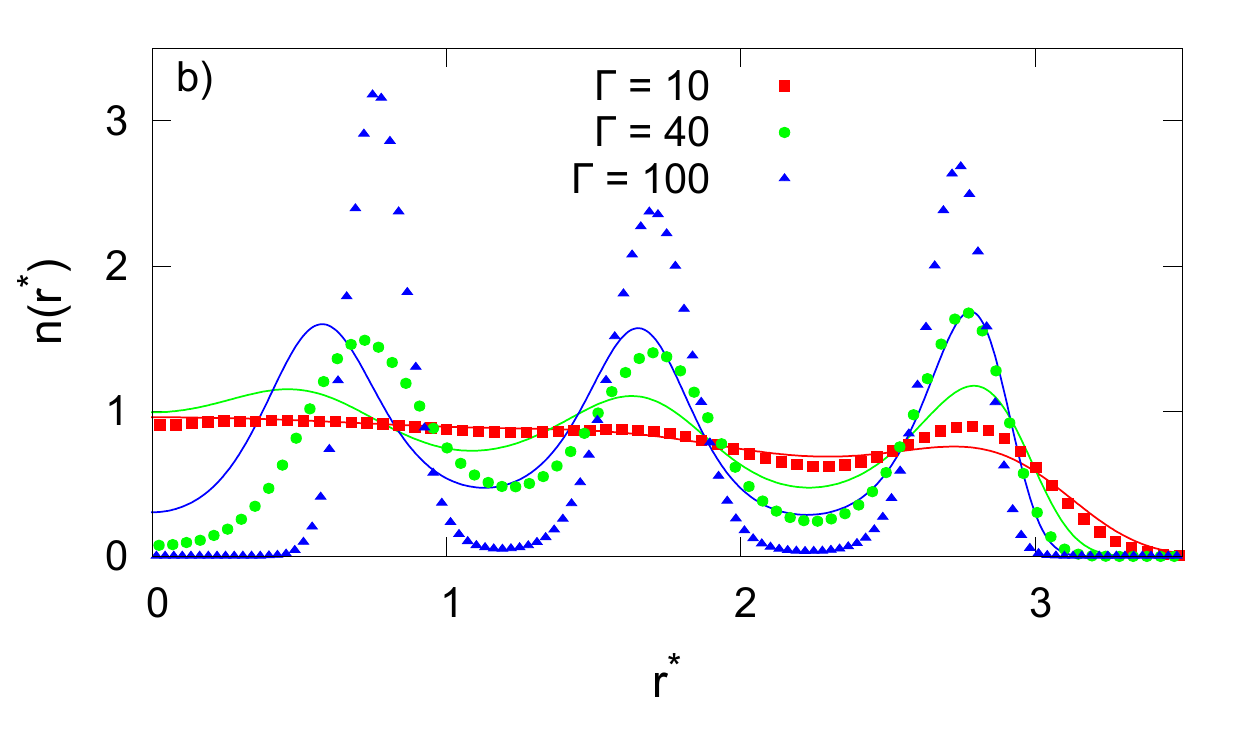}
\caption{Density profile for $N=100$ particles and various $\Gamma$ values: comparison of HNC results (solid lines) with Monte Carlo (symbols) for (a) Coulomb and (b) Yukawa interaction with $\protect\kappa^{\ast }=1$. }
\label{fig12}
\end{figure}

\begin{figure}[tbp]
\includegraphics[scale=0.6]{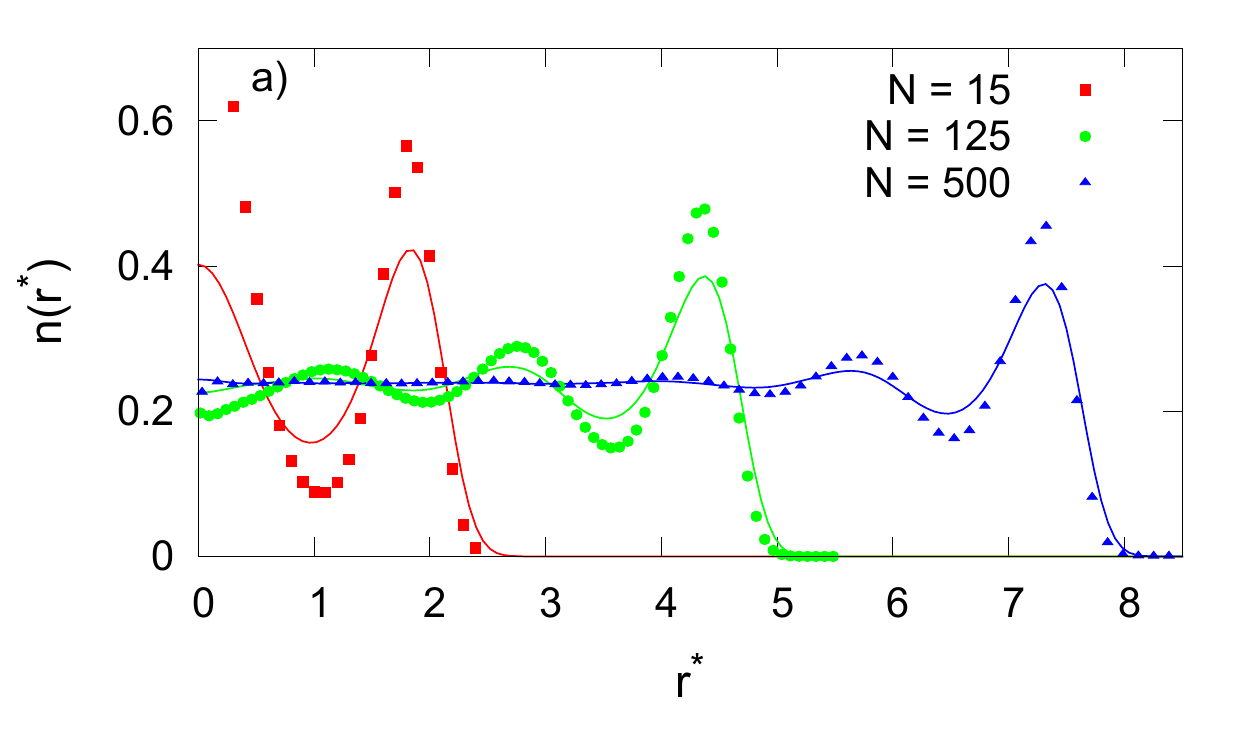}
\hfill \includegraphics[scale=0.6]{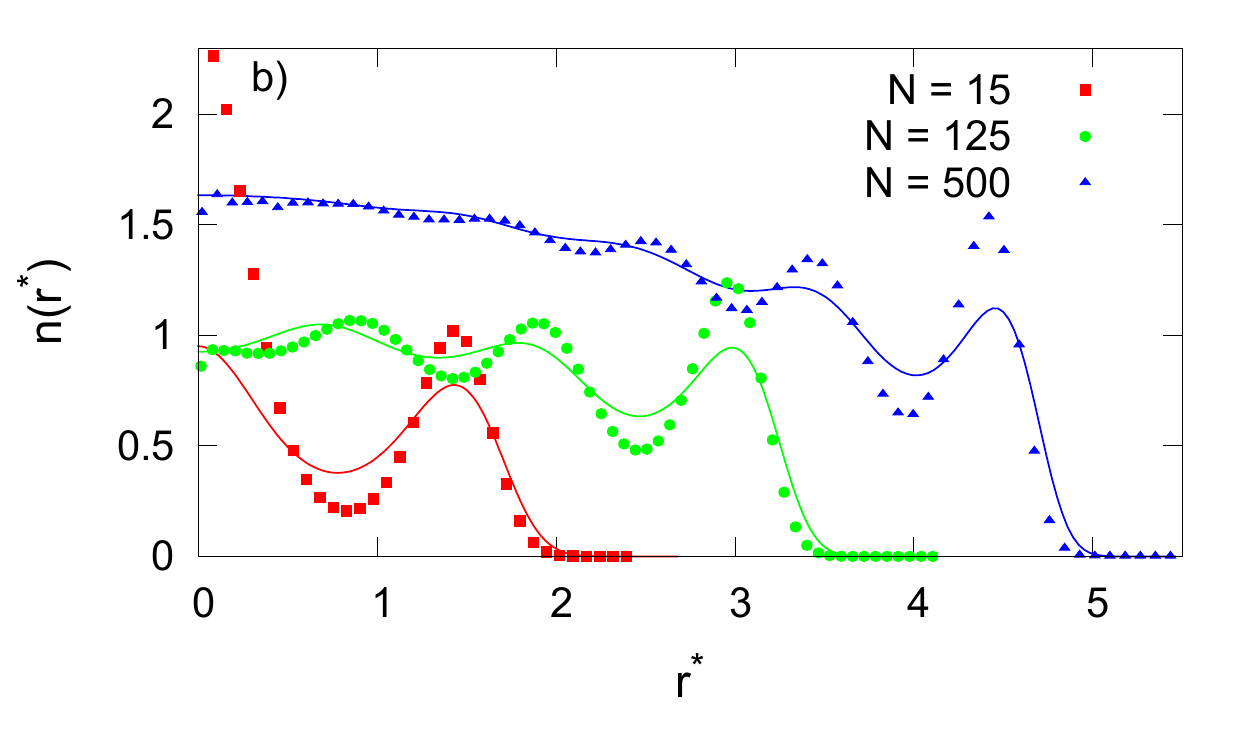}
\caption{Density profile for $\Gamma=20$: comparison of HNC results (lines) with Monte Carlo results (symbols) for (a) Coulomb and (b) Yukawa interaction with $\protect\kappa^{\ast }=1$.}
\label{fig13}
\end{figure}

The quality of HNC is tested by comparison to Monte Carlo simulations.
This is illustrated for $N=100$ and $\Gamma =10,40,100$
with $\kappa ^{\ast }=0$ in Fig. \ref{fig12}a) and with $\kappa ^{\ast }=1$ in Fig. \ref{fig12}b).
 HNC is a poor approximation at $r^{\ast }=0$ which results in overall poor
results for small particle numbers $\overline{N}<10$. This error appears
periodically with the creation of each new shell and is small if no particle
is at the center. For $\kappa ^{\ast }=0$ the shell locations match well the
simulation data, while increasing $\kappa ^{\ast }$ leads to decreased
accuracy for the inner shells. The effect is small up to $\kappa ^{\ast }=0.5
$. Figure \ref{fig13} compares HNC and Monte Carlo results with $N=15$, 125, and 500 for Coulomb charges in Fig. \ref{fig13}a), and for Yukawa charges with $\kappa^{\ast }=1$ in Fig. \ref{fig13}b). The shell populations (not  shown) and locations are nearly independent
of $\Gamma ,$ as seen in MC simulations \cite{MC1,MC2}, and are given
accurately by HNC. However, the peak heights and widths for the shells are
poorly predicted and require going beyond the HNC approximation.

\section{Adjusted HNC}
\label{AdHNC}

In a recent analysis for Coulomb systems in a spherical trap it was
also observed that the HNC approximation gives the correct location
and population of shells \cite{wrighton2009,wrighton2010}, which
depend only weakly on $\Gamma $. For Yukawa systems, these
properties become less accurate with increasing $\kappa ^{\ast }$.
For both Coulomb and Yukawa, the amplitude and width depend strongly
on $\Gamma $ and are underestimated by the HNC approximation at
strong coupling. This suggests that increasing the coupling constant
alone would increase the accuracy of HNC.

\subsection{Pair distribution function}

\begin{figure}[tbp]
\includegraphics[scale=0.6]{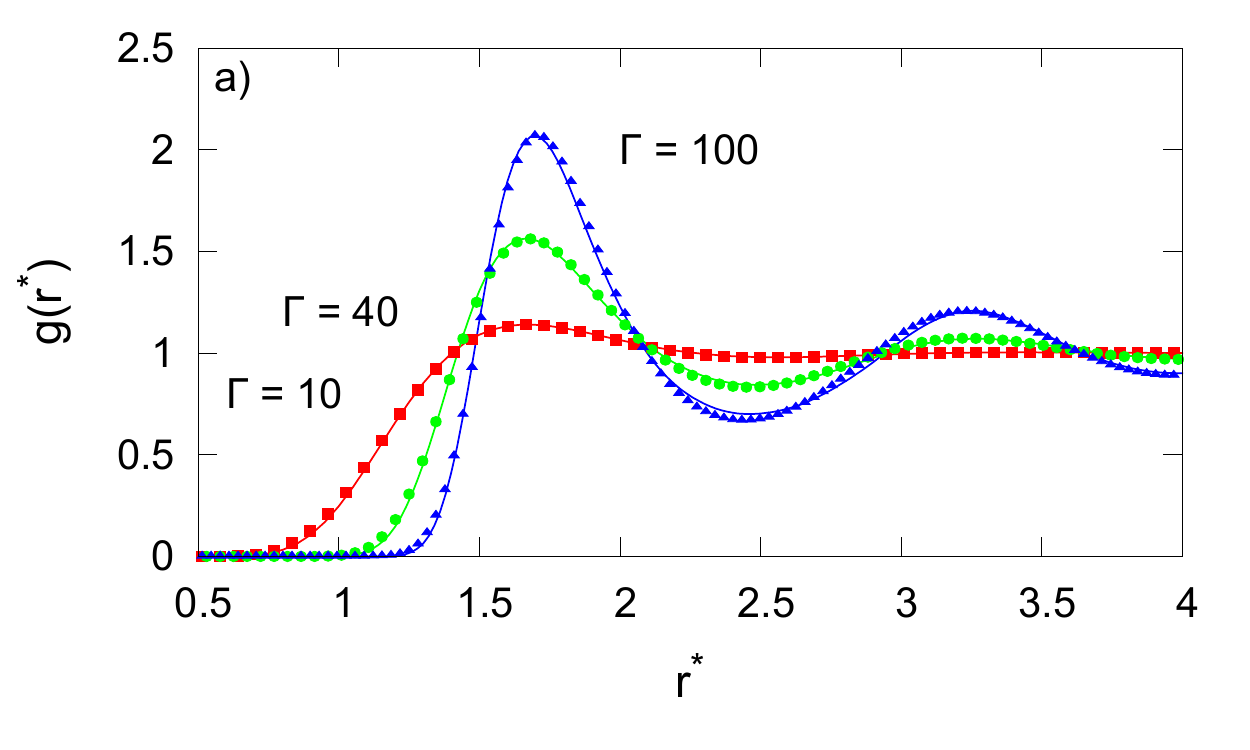} \hfill \includegraphics[scale=0.6]{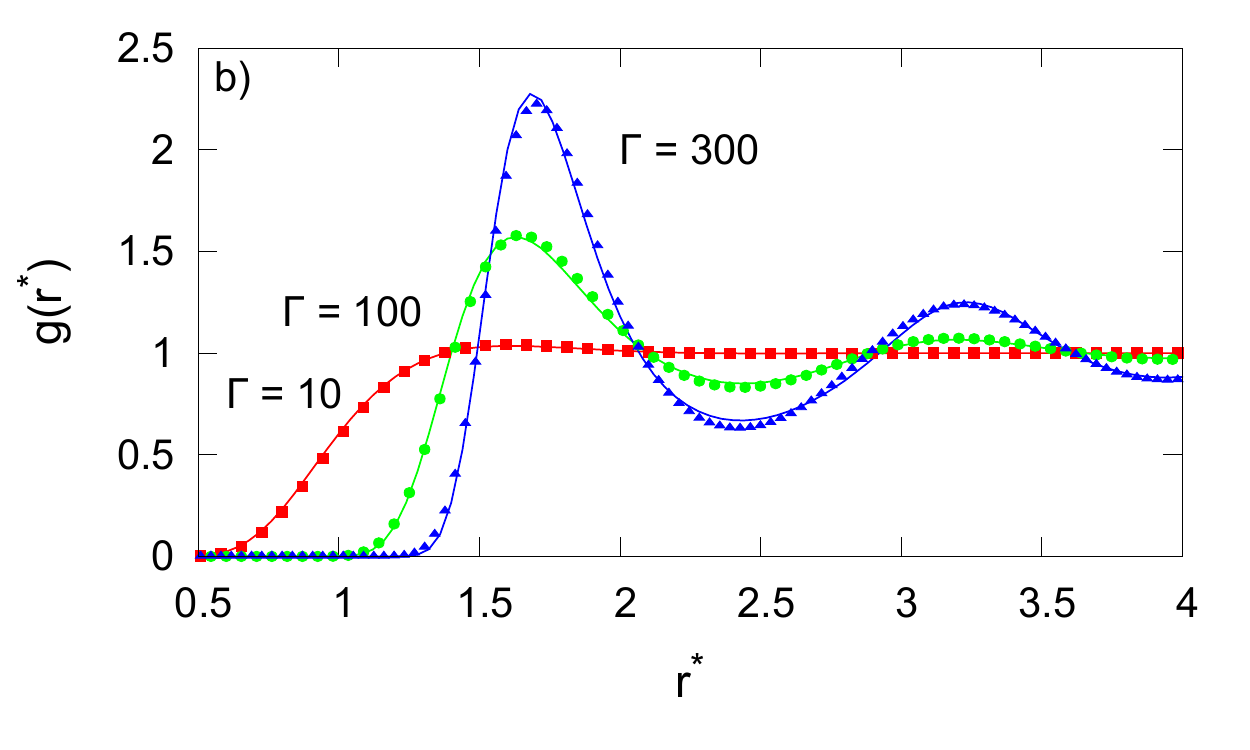}
\caption{Comparison of AHNC results for the pair distribution function $%
g(r^{\ast })$ with simulations for (a) Coulomb and (b) $\protect\kappa %
^{\ast }=2$. The values of $\Gamma ^{\prime }$ in (a) are $12.5$, $57$ and $%
160$, in (b) $10$, $130$ and $480$. } \label{fig13a}
\end{figure}
This failure of HNC for strong coupling has been studied in some
detail for the calculation of the Coulomb ${g_{\text{OCP}}(r)}$.
Among the earliest investigations is that of Ng \cite{NG} who
observed that the HNC peak positions are given accurately for strong
coupling, but not the amplitudes
and widths. He corrected the HNC by representing the bridge function of (\ref%
{3.5}) in the form
\begin{equation}
B_{\text{OCP}}(r^{\ast })\rightarrow \lambda (\Gamma )\beta
V(r^{\ast }), \label{4.1}
\end{equation}
where $\lambda (\Gamma )$ is a chosen function of $\Gamma $ and
$V(r^{\ast }) $ is the Coulomb potential. This particular choice was
not obtained from any theoretical analysis, but rather because it
leads back to the HNC form with a renormalized coupling constant
$\Gamma ^{\prime }=[1+\lambda (\Gamma )]\Gamma $. This approach will
be referred to as the adjusted HNC (AHNC). It was shown that an
accurate prediction of ${g_{\text{OCP}}(r)}$ could be
obtained over the entire fluid domain with the choice%
\begin{equation}
\lambda (\Gamma )\rightarrow \lambda _{Ng}(\Gamma )=0.6\erf
(0.024\Gamma ). \label{4.1a}
\end{equation}%
Subsequent theoretical studies of the Coulomb bridge function by
Rosenfeld and Ashcroft \cite{Rosenfeld}, indicated that it has a
"universal" form and hence could be represented by the corresponding
hard sphere bridge function for which an accurate parametrization
is known. Although considerably more complex to implement
computationally, it also gives a very good representation for
${g_{\text{OCP}}(r)}$. Furthermore, it has an important
thermodynamic consistency not shared by the HNC or AHNC
approximations. Evidently, the functional form (\ref{4.1})
represents the actual bridge
function for the relevant range of $r$ needed to determine ${g_{\text{OCP}}(r%
}^{\ast }{)}$ (the numerical difficulty of determining
$B_{\text{OCP}}\left( \mathbf{r}^{\ast }\right) $ precisely from
${g_{\text{OCP}}(r}^{\ast }{)}$ is discussed by Poll et al.
\cite{Poll}) Due to its simplicity and the direct interpretation as
a renormalization of the coupling strength the AHNC will be used here as
the means to improve the HNC approximation.

It remains to show how the bridge function should be chosen for the
Yukawa potential. An empirical choice has been suggested in the form
\cite{Daughton}
\begin{equation}
B_{\text{Y}}\left( \mathbf{r}^{\ast }\right) \rightarrow B_{\text{OCP}%
}\left( \mathbf{r}^{\ast }\right) e^{-\kappa ^{\ast 2}/2}.
\label{4.1b}
\end{equation}%
where $B_{\text{OCP}}\left( \mathbf{r}^{\ast }\right) $ is the
Coulomb
bridge function. This gives very good results for ${g_{\text{Y}}(r}^{\ast }{)%
}$ when $B_{\text{OCP}}\left( \mathbf{r}^{\ast }\right) $ is
approximated by the corresponding hard sphere bridge function, as
suggested by Rosenfeld and
Ashcroft. In contrast the AHNC for the Yukawa potential is obtained from (%
\ref{4.1})%
\begin{equation}
B_{\text{Y}}\left( \mathbf{r}^{\ast }\right) \rightarrow \lambda
(\Gamma )\beta \frac{e^{-\kappa r}}{r},
\end{equation}%
where $\lambda (\Gamma )$ is the same form as (\ref{4.1a}) for the OCP%
\begin{equation}
\lambda (\Gamma )\rightarrow \lambda _{Ng}(c(\kappa ^{\ast })\Gamma )=0.6%
\erf(c(\kappa ^{\ast })\Gamma ).
\label{4.1d}
\end{equation}%
The constant $c(\kappa ^{\ast })$ is adjusted for each $\kappa
^{\ast },$
with the Ng value $c(0)=0.024$. This Yukawa AHNC leads to the HNC form (\ref%
{3.9}), but with a renormalized coupling constant%
\begin{eqnarray}
&&\ln {g_{\text{AHNC}}(r^{\ast })}=-\Gamma ^{\prime
}\frac{e^{-\kappa ^{\ast
}r^{\ast }}}{r^{\ast }}+  \notag \\
&+&\overline{n}_{\text{T}}^{\ast }\int d\mathbf{r}^{\ast \prime }\left[ g_{%
\text{AHNC}}(r^{\ast \prime })-1\right] c_{\text{AHNC}}\left( |\mathbf{r}%
^{\ast }-\mathbf{r}^{\ast \prime }|\right) .
\end{eqnarray}%
Figures \ref{fig13a}a) and \ref{fig13a}b) show the excellent agreement 
between AHNC and molecular dynamics even at very strong
coupling for both $\kappa ^{\ast }=0$ and $\kappa ^{\ast }=2$. (Note
that $\Gamma $ and $\kappa ^{\ast }$ used here refer to a length
unit $r_{0}$ defined by Eq.~(\ref{2.4a}) with the OCP density). It
is interesting to note that results for recent MD simulations for
different values of $\Gamma $ and $\kappa ^{\ast }$ can be collapsed
in terms of a single effective coupling constant $\Gamma ^{\ast
}=\Gamma ^{\ast }\left( \Gamma ,\kappa ^{\ast }\right) $
\cite{Ott10}. In summary the AHNC for ${g(r^{\ast })}$ proposed by
Ng for the Coulomb OCP works as well for the strongly coupled Yukawa
plasma.


\subsection{Density profile}

With the results for the homogeneous OCP pair distribution function
as a
guide, a similar representation is considered for the bridge function $%
B(r|n) $ of the trap density profile (\ref{3.2})
\begin{equation}
B(r)\rightarrow \lambda (\Gamma )\beta V_{0}(r^{\ast }).
\label{4.3}
\end{equation}%
Here $\beta V_{0}$ is the trap potential $\Gamma r^{\ast 2}/2$,
restoring the HNC form (\ref{3.1}) and (\ref{3.8}) with a
renormalized $\Gamma ^{\prime }$. An initial approach would be to
use the same renormalization
function $\lambda (\Gamma )$ as obtained in the optimization of ${g_{\text{%
AHNC}}}$. This improves the accuracy for coupling constants up to
$\Gamma \simeq 40$. To include stronger coupling it is necessary to
choose a different renormalization function $\lambda (\Gamma )$ when
calculating the trap density profile. Although equations (\ref{4.1})
and (\ref{4.3})
formally allow for separate specifications of the renormalization function $%
\lambda(\Gamma)$ for the OCP and trap systems, results show that the
same function $\lambda(\Gamma)$ must be used to determine the trap
density profile to agree with simulations. That is,
$\lambda(\Gamma)$ can be determined by fitting either the density
profile or the pair correlation function, but not both.

The explanation as to why these two approaches (determining $\lambda
(\Gamma )$ separately for the systems and as a common function) give
very different results when using (\ref{4.1}) and (\ref{4.3}), lies
in the relationship between direct correlation functions at
different $\Gamma $. The scaled
direct correlation function (\ref{3.3}) is independent of $\Gamma $ for $%
\Gamma >10$. That is, if separate coupling constants for the trap ($\Gamma _{%
\mathrm{trap}}$) and OCP ($\Gamma _{\mathrm{OCP}}$) systems were
fitted, the direct correlation functions are still related by
\begin{equation}
\frac{c(r;\Gamma _{\mathrm{trap}})}{\Gamma _{\mathrm{trap}}}=\frac{%
c(r;\Gamma _{\mathrm{OCP}})}{\Gamma _{\mathrm{OCP}}}.
\end{equation}%
By considering again (\ref{3.2}), this shows that the two procedures of
using one common coupling constant, and using separate coupling
constants, are related by scaling the number of particles in the
trap. As the procedure of using a common renormalization function
has shown to give good results, an equivalent approach involving
separate renormalization functions is to have the effective number
of particles in the trap also be dependent on the coupling constant
so as to correct the discrepancy. It is important to note that with
this alternative approach, although both the coupling constant and
particle number are scaled, there still is only one fitting
parameter for the trap (which fixes both a scaled coupling constant
and particle number), and one fitting parameter for the OCP.
However, the interpretation of the scaled particle number for the
trap is not clear, as the shell structure depends critically on $N$.

Therefore we proceed by choosing to fit only the trap density
profile, and using a common renormalization function for both the
trap and the OCP systems.

\begin{figure}[tbp]
\includegraphics[scale=0.7]{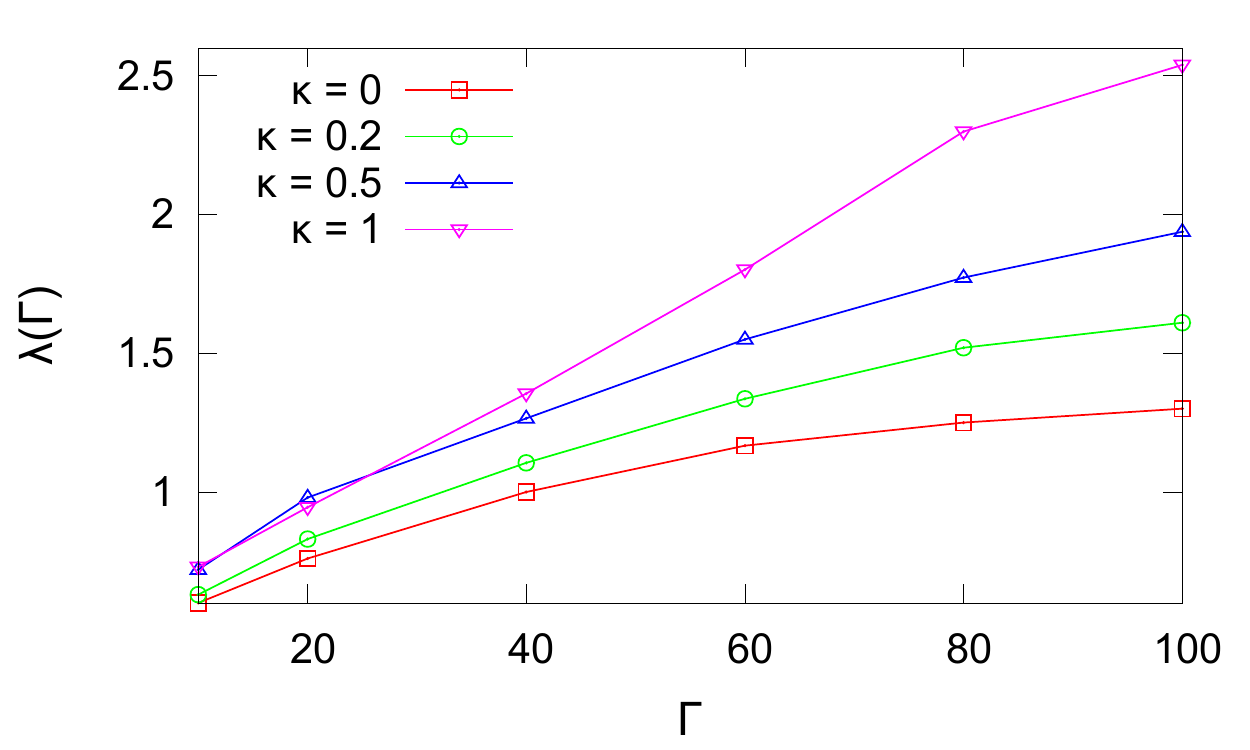}
\caption{Dependence of the AHNC parameter $\protect\lambda$ on $\Gamma$ and $%
\protect\kappa^{\ast}$ calculated by \eqref{4.4} at fixed particle number $%
N=100$. The renormalized coupling parameter is given by $\Gamma^{\prime }=[1+%
\protect\lambda(\Gamma)]\Gamma$.} \label{fig18}
\end{figure}

\begin{figure}[tbp]
\includegraphics[scale=0.7]{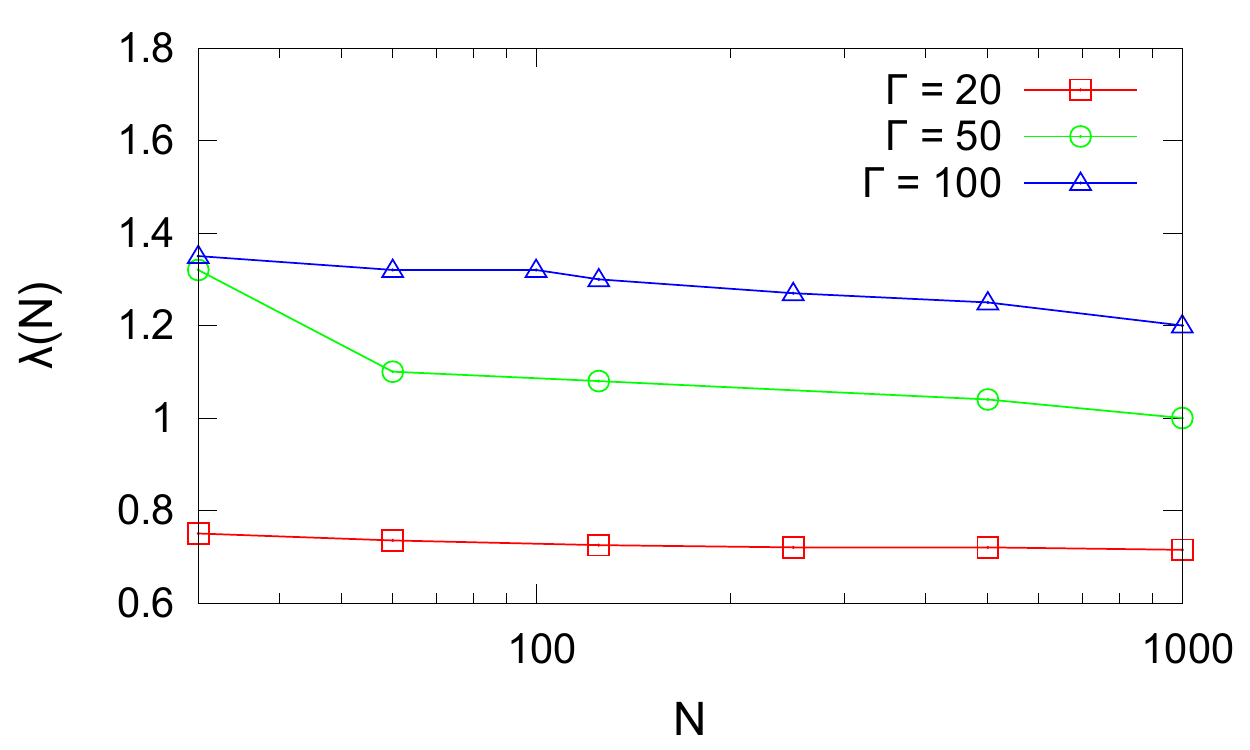}
\caption{Dependence of the AHNC parameter $\protect\lambda$ on $N$
for the
Coulomb case. The height of the outermost peak is used to obtain $\protect%
\lambda$. The renormalized coupling parameter is given by
$\Gamma^{\prime }=[1+\protect\lambda(\Gamma)]\Gamma$.} \label{fig19}
\end{figure}

An appropriate value of $\lambda (\Gamma )$ to optimize the density
profile
is obtained by minimizing the square difference of the Monte Carlo data $%
n_{MC}(r)$ and the HNC profile $n_{HNC}(r|\Gamma ^{\prime })$ with
respect to $\Gamma ^{\prime }$
\begin{equation}
\Gamma ^{\prime }:\min \int drr^{2}\left[ n_{\mathrm{MC}}(r|\Gamma )-n_{%
\mathrm{HNC}}(r|\Gamma ^{\prime })\right] ^{2}.  \label{4.4}
\end{equation}%
Since at small $r$ the HNC density profile is not accurate, the
difference in the peaks is weighted to the particle number in each
shell. In practice this effectively fits the height and width of the
outermost peak. The
dependence of $\lambda (\Gamma ,N)$ on $\Gamma $ and $N$ is shown in Fig. %
\ref{fig18} and Fig. \ref{fig19}. The $\Gamma $ dependence for different $%
\kappa ^{\ast }$ cannot be collapsed to a single curve by rescaling
$\Gamma $
as in (\ref{4.1d}), as the asymptotic large $\kappa ^{\ast }$ limits of $%
\lambda (\Gamma ,N)$ are now different.

\begin{figure}[tbp]
\includegraphics[scale=0.7]{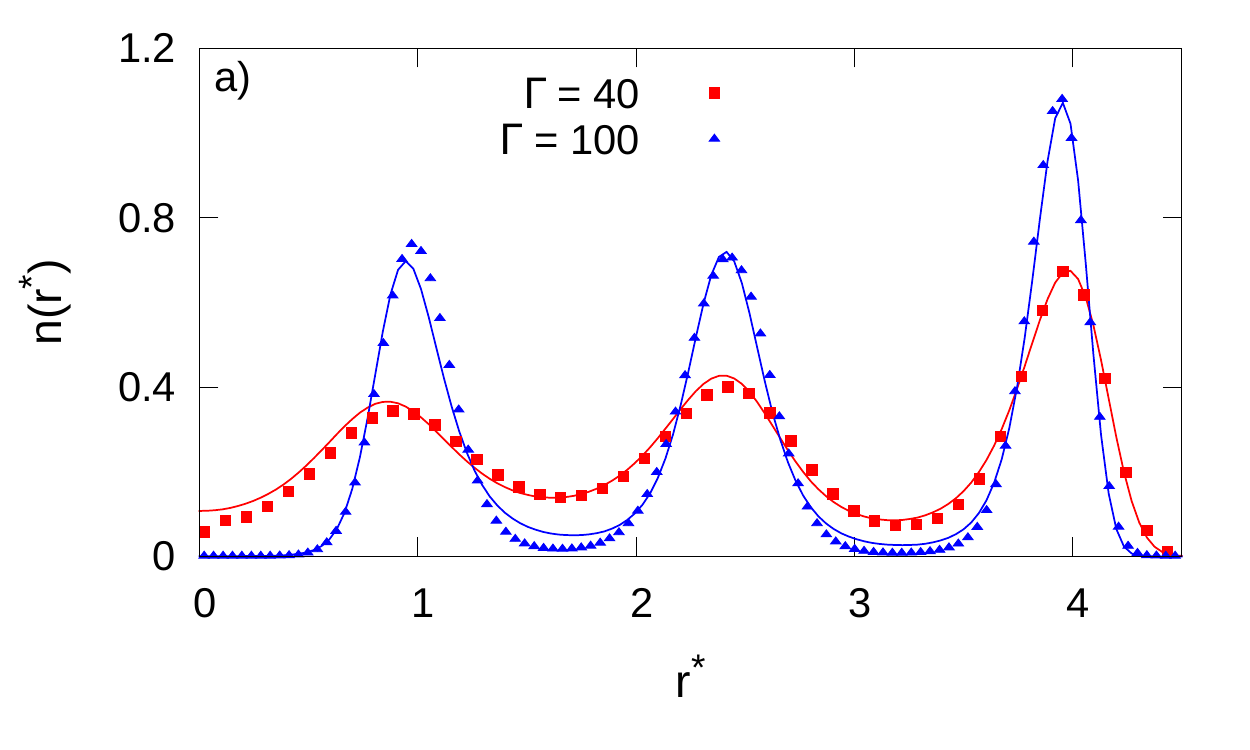} \hfill \includegraphics[scale=0.7]{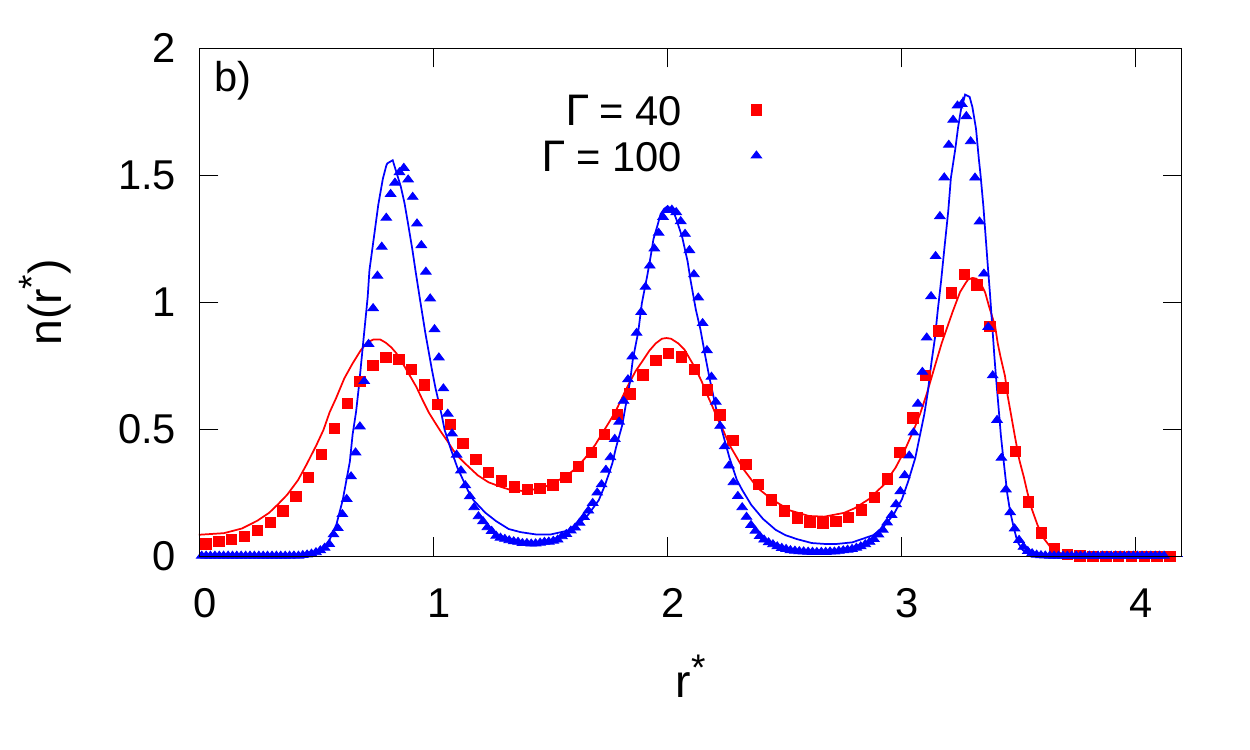}
\hfill \includegraphics[scale=0.7]{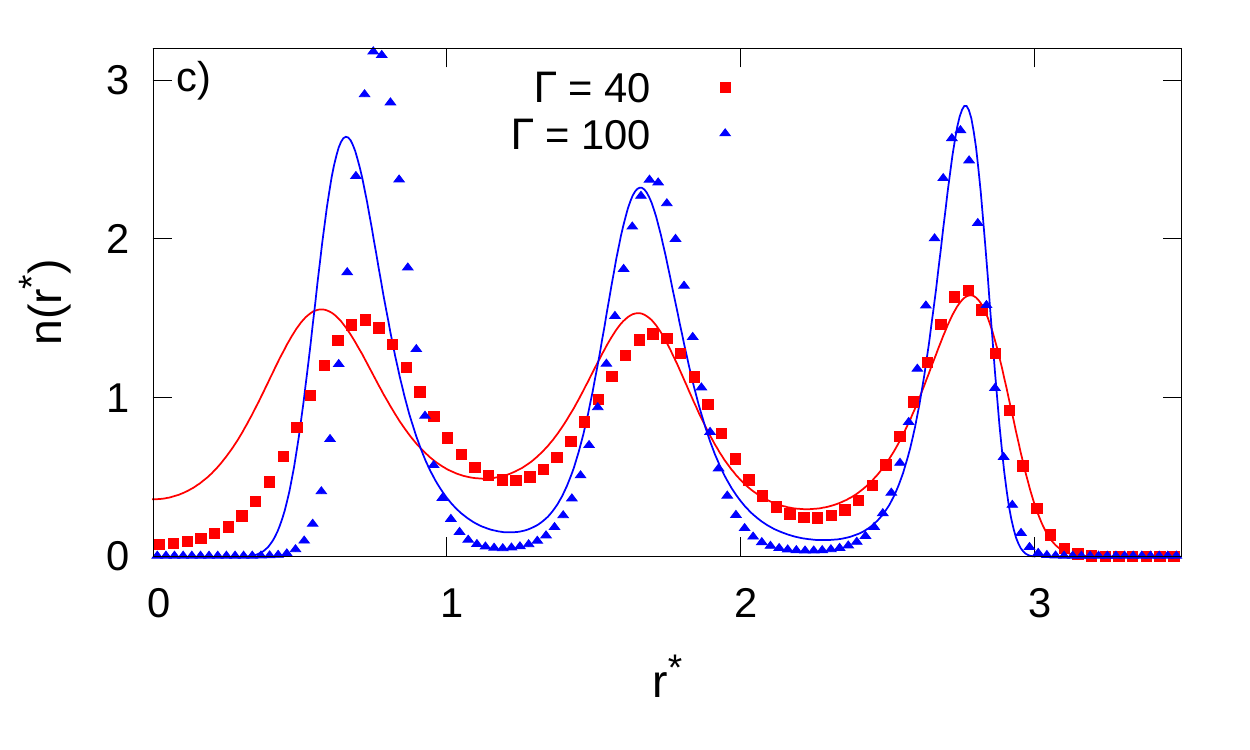}
\caption{Density profile for $N=100$ particles: Comparison of AHNC
results (lines) and MC data (symbols) for (a) Coulomb interaction, and Yukawa interaction 
with (b) $\protect\kappa ^{\ast }=0.5$ and (c) $\protect\kappa ^{\ast }=1$. The 
renormalized $\Gamma ^{\prime }$ values used are shown in Fig.
\protect\ref{fig18}.} \label{fig15}
\end{figure}

\begin{figure}[tbp]
\includegraphics[scale=0.7]{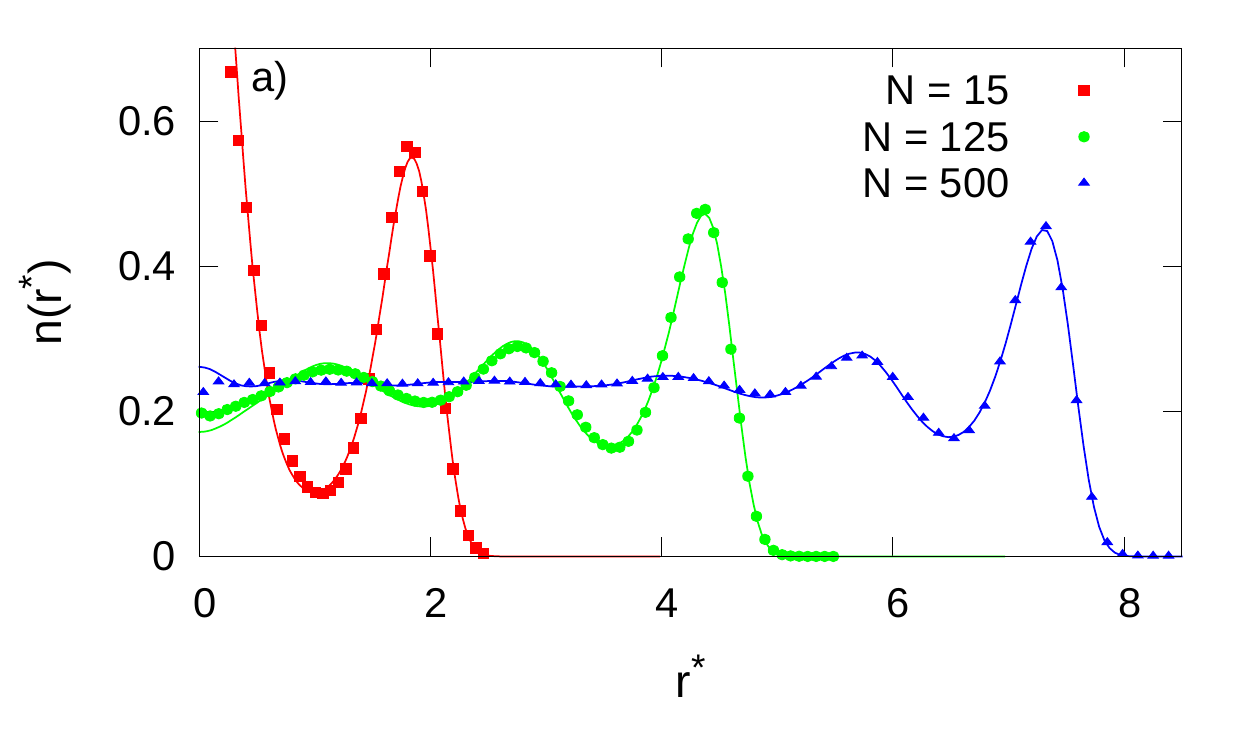} \hfill \includegraphics[scale=0.7]{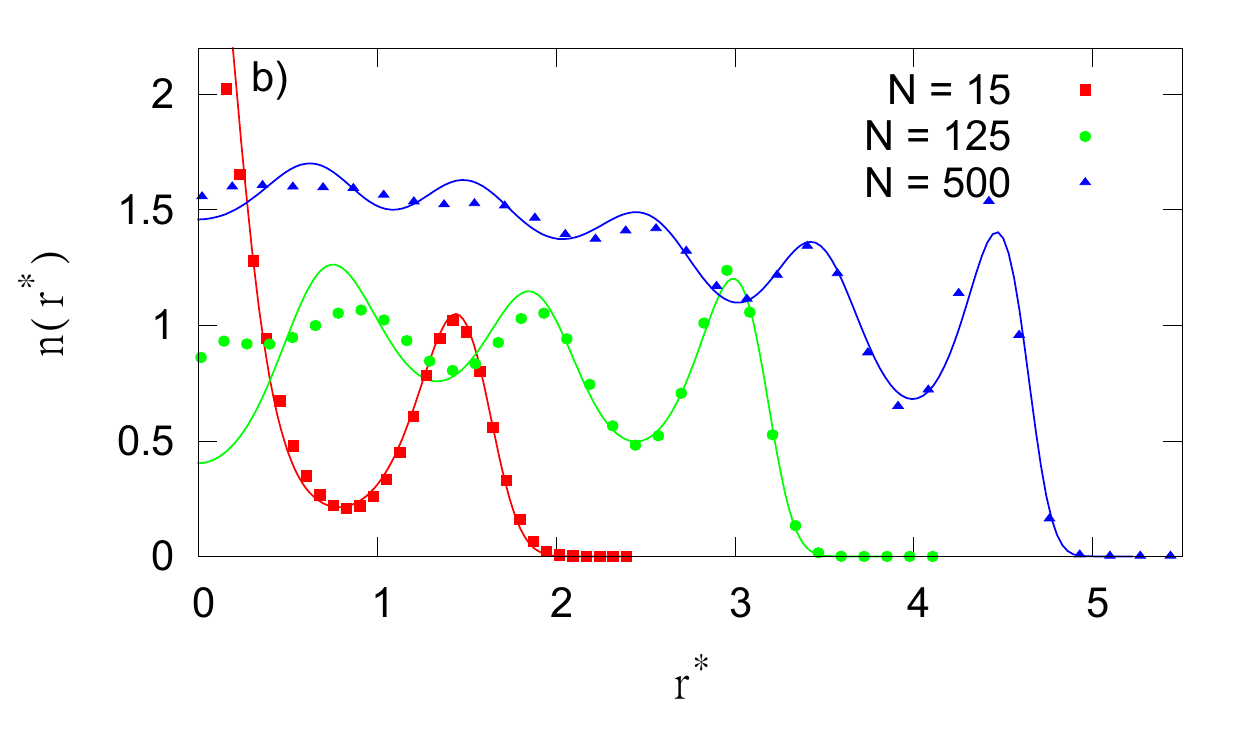}
\caption{Density profile for a fixed coupling constant $\Gamma=20$
and various particle numbers: comparison of AHNC results (lines) and
MC data
(symbols) for (a) a Coulomb and (b) a Yukawa system with $\protect\kappa%
^{\ast}=1$ . Disagreement in the inner part is increasing with
particle number for the Yukawa system. } \label{fig17}
\end{figure}

The quality of the AHNC approximation is again established by
comparison to Monte Carlo data, cf. Fig. \ref{fig15}. AHNC describes
accurately the
density profile of Coulomb charges for the full range of $\Gamma $ [Fig. \ref%
{fig15} (a)] and particle numbers $N$ [Fig. \ref{fig17} (a)], while
keeping the simple form of HNC. A similar improvement in accuracy is
observed for the Yukawa system with $\kappa ^{\ast }\leq 0.5$ and
$N<100$ [Fig. \ref{fig15} (b)]. For
larger $\kappa ^{\ast }$ errors in the inner shells occur and
increase with increasing $\kappa ^{\ast }$ and $N$ [Fig. \ref{fig15}
(c) and Fig. \ref{fig17} (b)]. For large particle numbers fitting
$\lambda $ becomes more subjective, depending upon which criteria
are imposed, e.g. best outermost peak height or inner shell heights.
Similarly, increasing the renormalized coupling constant beyond a
certain value is only trading agreement from inner to outer shells.

It is curious that the AHNC procedure works so well for pair
correlations, both Coulomb and Yukawa, and for the Coulomb trap
density profile, but fails for the Yukawa density profile at large
$\kappa ^{\ast }$ and $N$. One possible explanation is the
following. The density equation of the HNC entails an additional
approximation not contained in that for the pair correlations,
namely that the pair correlations in the trap can be represented by
those of the OCP. This can be justified for Coulomb interactions,
but that argument does not extend to Yukawa interactions. At large
$\kappa ^{\ast }$ this approximation may no longer hold. In
addition, the shell structure is enhanced at large $\kappa ^{\ast }$,
and the number of shells increases with $N$. Hence there are
increased demands on the AHNC to represent more complex structure.

There is a qualitative difference between the Coulomb and Yukawa
cases at large $N$. In the former case, the harmonic trap is exactly
equal to the effect of a uniform neutralizing background, and the
system approaches the Coulomb OCP for large $N$ except at the
boundaries. However, for the Yukawa case the relationship of the
trap to the neutralizing background no longer holds. Totsuji et al. 
have derived the corresponding confinement potential for a Yukawa 
system~\cite{Totsuji05a,Totsuji05b}.

\section{Discussion}

A theoretical description is developed for the shell structure of
spherically confined Yukawa plasmas. While the precise shell
occupations are
well known from computer simulations, both for trapped Coulomb, e.g. \cite%
{rafac_pnas91,ludwig05} and Yukawa plasmas e.g. \cite{e}, it is
desirable to have an analytical theory that correctly reproduces
these results and provides physical insight into the correlation
properties. Classical density functional theory is the proper
starting point for this. In particular, it has been shown that the HNC
approximation is able to provide the density profile (the formation,
shape, location, and population of shells) accurately for weak to
moderate coupling ($\Gamma <10$). However, HNC fails to reproduce
the correct width of the shells.

A simple representation of the bridge functions $B$ (corrections to
HNC) called adjusted HNC (AHNC) is able to provide quantitative
agreement in the case of Coulomb interactions for $\Gamma \leq 100$
and $N\leq 500$, indicating that a simple renormalization of the HNC
is sufficient to capture the structural effects of confinement. A
similar adjusted HNC provides substantial improvement for the
isotropically trapped Yukawa system as well. While it correctly
reproduces the shape of the outermost shell(s) that host the
majority of particles, it is less accurate for the inner shells, in
particular with increasing screening parameter and particle number.

\section{Acknowledgments}

This work is supported by the Deutsche Forschungsgemeinschaft via
SFB-TR 24, and by the NSF/DOE Partnership in Basic Plasma Science
and Engineering under Department of Energy Award No. 
DE-FG02-07ER54946.



{}

\end{document}